\documentclass[aps,showpacs,twocolumn]{revtex4}
\usepackage{amsmath}
\usepackage{epsfig}
\usepackage{color}

\begin{document}

\title{Theoretical Study of a $d^{*}$ resonance in the coupled $^{3}D_{3}$ - $^{3}G_{3}$ partial
waves of nucleon-nucleon scattering}

\author{Hongxia Huang$^a$, Jialun Ping$^a$\footnote{jlping@njnu.edu.cn, corresponding author},
        Chengrong Deng$^b$ and Fan Wang$^c$}

\affiliation{$^a$Department of Physics, and Jiangsu Key
Laboratory for Numerical Simulation of Large Scale Complex Systems,
Nanjing Normal University, Nanjing 210023, China}
\affiliation{$^b$School of Mathematics and Physics, Chongqing Jiaotong University, Chongqing 400074, China}
\affiliation{$^c$Department of Physics, Nanjing University, Nanjing 210093, China}

\begin{abstract}
We calculated the
phase shifts of the coupled $^{3}D_{3}$ - $^{3}G_{3}$
partial waves of nucleon-nucleon scattering taking into account the
$^7S_3$ $\Delta-\Delta$ channel coupling in the framework of
two constituent quark models: the quark delocalization color
screening model and the chiral quark model. Our results show that
there is a resonance due to $^{7}S_{3}^{\Delta\Delta}$ coupling to the
$^{3}D_{3}^{NN}$ and $^{3}G_{3}^{NN}$ partial waves in both of
these two models, which is consistent with the d* resonance observed
by WASA-at-COSY collaboration. The resonance structure in the
$^{3}D_{3}^{NN}$ partial wave is remarkable, whereas in the
$^{3}G_{3}^{NN}$ phase shifts there is only a small bump around the
resonance energy. This result is in agreement with the recent
experimental results of WASA-at-COSY Collaboration.
\end{abstract}

\pacs{13.75.Cs, 12.39.Pn, 12.39.Jh, 14.20.Pt}

\maketitle

\section{\label{sec:introduction}Introduction}
Recently, a narrow resonance-like structure has been observed in
the reaction $pn \rightarrow d\pi^{0}\pi^{0}$ and $pn \rightarrow
d\pi^{+}\pi^{-}$ at a mass $M = 2.37$ GeV with width $\Gamma
\approx 70$ MeV~\cite{ABC1,ABC2,ABC3}, which indicates the
existence of an $IJ^{P} = 03^{+}$ sub-threshold $\Delta\Delta$
resonance, called $d^{*}$ in the literature. Additional evidences
had been found in the exclusive measurements of the
quasi-free $pn \rightarrow pp\pi^{0}\pi^{-}$ reaction~\cite{ABC4}.
A. Pricking $et~al.$ suggested that if this observed resonance
structure constitutes an $s-$channel resonance in the $pn$ system, then it
should cause distinctive consequences in $pn$
scattering~\cite{Annette}. Then they added the resonance amplitude
in the $^{3}D_{3}$ ($^{3}G_{3}$) partial wave analysis, estimated
the resonance effect in $pn$ scattering and found that the
resonance effect improves considerably the description of total
cross section of $pn$ scattering and the vector analyzing power
$A_{y}$ exhibits the largest sensitivity to the resonance.
High precision data are needed for the energy region
$T_{n}=1.0-1.3$ GeV to have a crucial test of the resonance
hypothesis. Such measurements have actually been carried out very
recently with the WASA detector at COSY, in which the $pn$
analyzing power $A_{y}$ was measured over a large angular
range~\cite{ABC5}. Incorporating the new $A_{y}$ data into the
SAID analysis produces a pole in the $^{3}D_{3}$ - $^{3}G_{3}$
partial waves as expected from the $d^{*}$ resonance hypothesis.

The possibility of $IJ^P=03^+$ dibaryon state was first proposed by Dyson and
Xuong in 1964~\cite{Dyson}, in which a bound $\Delta\Delta$ state was
predicted to lie at $2350$ MeV. Goldman $et~al.$~\cite{Goldman} showed that the $d^{*}$
should be bound in any QCD based quark model that incorporates color confinement, hyperfine
color-magnetic interaction and quark delocalization. They called it
inevitable nonstrange dibaryon d*. The recent
three-body Faddeev equation calculation supports the existence of
$d^{*}$~\cite{faddeev}. M. Bashkanov $et~al.$ further pointed out
that the observation of the $d^{*}$ resonance state will rewarm
the search of other novel six-quark
configurations allowed by QCD~\cite{Bashkanov}. Our former calculations
showed that the $I=0$, $J=3$ $d^{*}$ is a tightly bound six-quark
system rather than a loosely bound nucleus-like system of two
$\Delta$s~\cite{QDCSM0,QDCSM1,QDCSM2}. Recently, our group
calculated the resonance mass and decay width of the
$IJ^{P}=03^{+}$ and $IJ^{P}=30^{+}$ $\Delta\Delta$ states and
showed that the $IJ^{P}=03^{+}$ $\Delta\Delta$ resonance is a
promising candidate for the recent observed one in the so-called
ABC-effect~\cite{Ping_NN,Huang}. However, where we only calculate
the $d^{*}$ resonance in the $^{3}D_{3}$ partial wave of nucleon-nucleon
($NN$) scattering~\cite{Ping_NN}. Whether there is a resonance
structure in the coupled $^{3}D_{3}$ - $^{3}G_{3}$ partial waves
of nucleon-nucleon scattering as expected in Ref.~\cite{Annette}
and the experimental observations of Ref.~\cite{ABC5}?
This article reports the results of a channel
coupling calculation of the $I=0$ $NN$ scattering in the coupled
$^{3}D_{3}$ - $^{3}G_{3}$ partial waves with and without the
$^7S_3$ $d^*$ structure.

The direct use of Quantum chromodynamics (QCD), the fundamental
theory of the strong interaction, in nucleon-nucleon interaction
is still out of reach of the present techniques, although the
lattice QCD has made a considerable progress recently~\cite{LQCD}.
QCD-inspired quark models
are still the main approach to study the baryon-baryon
interaction. In our former study of the nucleon-nucleon scattering
and the $d^{*}$ resonance state, two quark models were used:
one is the chiral quark model (ChQM)~\cite{ChQM1}, in which
the $\sigma$ meson is indispensable to provide the
intermediate-range attraction;
The other is the quark delocalization color screening
model (QDCSM)~\cite{QDCSM0}, which has been developed with the aim
to understand the well-known similarities between nuclear and
molecular forces despite the obvious energy and length scale
differences. In this model, the intermediate-range attraction is
achieved by the quark delocalization, which is like the electron
percolation in the molecules. The color screening is needed to
make the quark delocalization feasible and it might be an
effective description of the hidden color channel
coupling~\cite{Huang2}. Both two models have been successfully
applied to hadron spectroscopy and $NN$ interaction.
In this work, we
use these two quark models to study the resonance structure in the
$^{3}D_{3}$, $^{3}G_{3}$ and the coupled $^{3}D_{3}$ - $^{3}G_{3}$
partial waves of $NN$ scattering.

The structure of this paper is as follows. A brief introduction of
two quark models is given in section II. Section III devotes to
the numerical results and discussions. The last section is a
summary.

\section{Two quark models}
The two quark models: chiral quark model and quark delocalization color
screening model, have been used in our previous work to study the
$NN$ N-hyperon interaction and dibaryon~\cite{Ping_NN}. The details of two models can be found in
Refs.~\cite{ChQM1,QDCSM0,QDCSM1,Ping_NN}. In the following, only the
Hamiltonians and parameters are given.

\subsection{Chiral quark  model}

The ChQM Hamiltonian in $NN$ sector is
\begin{eqnarray}
H &=& \sum_{i=1}^6 \left(m_i+\frac{p_i^2}{2m_i}\right) -T_c
\nonumber \\
&+ & \sum_{i<j} \left[
V^{G}(r_{ij})+V^{\pi}(r_{ij})+V^{\sigma}(r_{ij})+V^{C}(r_{ij})
\right], \nonumber \\
V^{G}(r_{ij})&=& \frac{1}{4}\alpha_s {\mathbf \lambda}_i \cdot
{\mathbf \lambda}_j
\left[\frac{1}{r_{ij}}-\frac{\pi}{m_q^2}\left(1+\frac{2}{3}
{\mathbf \sigma}_i\cdot {\mathbf\sigma}_j \right)
\delta(r_{ij}) \right. \nonumber \\
& & \left. -\frac{3}{4m_q^2r^3_{ij}}S_{ij}\right]+V^{G,LS}_{ij},
\nonumber \\
V^{G,LS}_{ij} & = & -\frac{\alpha_s}{4}{\mathbf \lambda}_i
\cdot{\mathbf \lambda}_j
\frac{1}{8m_q^2}\frac{3}{r_{ij}^3}[{\mathbf r}_{ij} \times
({\mathbf p}_i-{\mathbf p}_j)] \cdot({\mathbf \sigma}_i+{\mathbf
\sigma}_j),
\nonumber \\
V^{\pi}(r_{ij})&=& \frac{1}{3}\alpha_{ch}
\frac{\Lambda^2}{\Lambda^2-m_{\pi}^2}m_\pi \left\{ \left[ Y(m_\pi
r_{ij})- \frac{\Lambda^3}{m_{\pi}^3}Y(\Lambda r_{ij}) \right]
 \right.\nonumber \\
&& \!\!\! {\mathbf \sigma}_i \cdot{\mathbf \sigma}_j
\left. +\left[ H(m_\pi r_{ij})-\frac{\Lambda^3}{m_\pi^3}
H(\Lambda r_{ij})\right] S_{ij} \right\} {\mathbf \tau}_i
\cdot {\mathbf \tau}_j, \nonumber  \\
V^{\sigma}(r_{ij})&=& -\alpha_{ch} \frac{4m_u^2}{m_\pi^2}
\frac{\Lambda^2m_\sigma}{\Lambda^2-m_{\sigma}^2} \left[ Y(m_\sigma
r_{ij})-\frac{\Lambda}{m_\sigma}Y(\Lambda r_{ij})
\right]  \nonumber \\
& + & V^{\sigma,LS}_{ij}, \\
V^{\sigma,LS}_{ij} & = & -\frac{\alpha_{ch}}{2m_{\pi}^2}
\frac{\Lambda^2}{\Lambda^2-m_{\sigma}^2}m^3_{\sigma} \left[
G(m_\sigma r_{ij})- \frac{\Lambda^3}{m_{\sigma}^3}G(\Lambda
r_{ij}) \right]
\nonumber \\
& & [{\mathbf r}_{ij} \times ({\mathbf p}_i-{\mathbf
p}_j)] \cdot({\mathbf \sigma}_i+{\mathbf \sigma}_j), \nonumber
\end{eqnarray}
\begin{eqnarray}
V^{C}(r_{ij})&=& -a_c {\mathbf \lambda}_i \cdot {\mathbf
\lambda}_j (r^2_{ij}+V_0)+V^{C,LS}_{ij}, \nonumber \\
V^{C,LS}_{ij} & = & -a_c {\mathbf \lambda}_i \cdot{\mathbf
\lambda}_j
\frac{1}{8m_q^2}\frac{1}{r_{ij}}\frac{df(r_{ij})}{dr_{ij}}[{\mathbf
r}_{ij} \times ({\mathbf p}_i-{\mathbf p}_j)]
\nonumber \\
& &  \cdot ({\mathbf
\sigma}_i+{\mathbf \sigma}_j),~~~~~~ f(r_{ij})=r^{2}_{ij},
\nonumber \\
S_{ij} & = &  \frac{{\mathbf (\sigma}_i \cdot {\mathbf r}_{ij})
({\mathbf \sigma}_j \cdot {\mathbf
r}_{ij})}{r_{ij}^2}-\frac{1}{3}~{\mathbf \sigma}_i \cdot {\mathbf
\sigma}_j. \nonumber
\end{eqnarray}
Where $S_{ij}$ is quark tensor operator, $Y(x)$, $H(x)$ and $G(x)$
are standard Yukawa functions, $T_c$ is the kinetic energy of the
center of mass. All other symbols have their usual meanings.

\subsection{Quark delocalization color screening model}

The Hamiltonian of QDCSM has the same form as Eq.(1), but without
$\sigma$ meson exchange and a phenomenological color screening confinement
potential is used,
\begin{eqnarray}
V^{C}(r_{ij})&=& -a_c {\mathbf \lambda}_i \cdot {\mathbf
\lambda}_j [f(r_{ij})+V_0]+V^{C,LS}_{ij}, \nonumber
\\
 f(r_{ij}) & = &  \left\{ \begin{array}{ll}
 r_{ij}^2 &
 \qquad \mbox{if }i,j\mbox{ occur in the same } \\
 & \qquad \mbox{baryon orbit}, \\
 \frac{1 - e^{-\mu r_{ij}^2} }{\mu} & \qquad
 \mbox{if }i,j\mbox{ occur in different} \\
 & \qquad \mbox{baryon orbits}.
 \end{array} \right.
\end{eqnarray}
Here, $\mu$ is the color screening constant to be determined by
fitting the deuteron mass. The quark delocalization
in QDCSM is realized by
replacing the left- and right-centered Gaussian functions,
the single-particle orbital wave functions in the usual quark
cluster model,
\begin{eqnarray}
\phi_{\alpha}(\vec{S}_i) & = & \left( \frac{1}{\pi b^2}
\right)^{3/4}
   e^{-\frac{1}{2b^2} (\vec{r}_{\alpha} - \vec{S}_i/2)^2}  \\
\phi_{\beta}(-\vec{S}_i) & = & \left( \frac{1}{\pi b^2}
\right)^{3/4}
   e^{-\frac{1}{2b^2} (\vec{r}_{\beta} + \vec{S}_i/2)^2}.
\end{eqnarray}
with delocalized ones,
\begin{eqnarray}
\psi_{\alpha}(\vec{S}_i ,\epsilon) & = & \left(
\phi_{\alpha}(\vec{S}_i)
+ \epsilon \phi_{\alpha}(-\vec{S}_i)\right) /N(\epsilon), \nonumber \\
\psi_{\beta}(-\vec{S}_i ,\epsilon) & = &
\left(\phi_{\beta}(-\vec{S}_i)
+ \epsilon \phi_{\beta}(\vec{S}_i)\right) /N(\epsilon), \label{1q} \\
N(\epsilon) & = & \sqrt{1+\epsilon^2+2\epsilon e^{-S_i^2/4b^2}}.
\nonumber
\end{eqnarray}

The parameters of two models are fixed by baryon and deuteron properties
and/or nucleon-nucleon scattering phase shifts, they are listed in
Table \ref{parameters}.

\begin{table}[ht]
\caption{Parameters of quark models}
\begin{tabular}{lcc}
\hline\hline
 & {\rm ChQM} & ~~~~{\rm QDCSM}     \\
\hline
$m_{u,d}({\rm MeV})$        &  313    &  ~~~~313     \\
$b ({\rm fm})$              &  0.518  &  ~~~~0.518   \\
$a_c({\rm MeV\,fm}^{-2})$   &  46.938 & ~~~~56.755   \\
$V_0({\rm fm}^{2})$         &  -1.297 & ~~~~-0.5279 \\
$\mu ({\rm fm}^{-2})$       &   --     &  ~~~~0.45        \\
$\alpha_s$                  &  0.485  &  ~~~~0.485  \\
$m_\pi({\rm MeV})$          &  138    &  ~~~~138      \\
$\alpha_{ch}$               &  0.027  & ~~~~0.027   \\
$m_\sigma ({\rm MeV})$      &  675    &    ~~~~--          \\
$\Lambda ({\rm fm}^{-1})$   &  4.2    &  ~~~~4.2     \\
\hline\hline
\end{tabular}
\label{parameters}
\end{table}

\subsection{The calculation method}
To calculate the $NN$ scattering phase shifts and
resonance states, the well developed resonating group
method (RGM) is used. The details of RGM can be found in
Ref.\cite{RGM}. Here only the necessary equations are given. In
RGM, the multiquark wave
functions are approximated by the cluster wave
functions, the internal motions of clusters are frozen and the
relative motion wave-function $\chi(\boldsymbol{R})$ satisfies the
following RGM equation
\begin{equation}
\int H(\boldsymbol{R''}, \boldsymbol{R'}) \chi (\boldsymbol{R'})
d\boldsymbol{R'} = E \int N(\boldsymbol{R''},\boldsymbol{R'}) \chi
(\boldsymbol{R'}) d\boldsymbol{R'} , \label{RGM-1}
\end{equation}
where
\begin{eqnarray}
\left\{ \begin{array}{c}
         H(\boldsymbol{R'',R'}) \\
         N(\boldsymbol{R'',R'}) \\
        \end{array} \right\} & = & \left\langle
{\cal A}[\phi_1\phi_2\delta(\boldsymbol{R-R''})] \right\vert
\left\{\begin{array}{c}
         H \\
         1 \\
         \end{array} \right\} \nonumber \\
 & & \left\vert
{\cal A}[\phi_1\phi_2\delta(\boldsymbol{R-R'})] \right\rangle .
\label{RGM-2}
\end{eqnarray}
where $\phi_1,\phi_2$ are the internal wavefunctions of two clusters.
${\cal A}=1+{\cal A}'$ is the anti-symmetrization operator and has
the properties, ${\cal A}H=H{\cal A}$, ${\cal A}^2={\cal A}$. With
this properties, the RGM equation can be written as an
integro-differential equation
\begin{eqnarray}
\left[
\frac{\nabla^2_{\boldsymbol{R''}}}{2\mu}-{V}^{D}(\boldsymbol{R''})+E_{CM}
 \right]  \chi(\boldsymbol{R''}) & = & \nonumber \\
& & \hspace{-1.1in} \int
W_{L}(\boldsymbol{R'',R'})\chi(\boldsymbol{R'})d\boldsymbol{R'},
\label{RGM-4}
\end{eqnarray}
where $E_{CM}=E-E_{int}$ is the kinetic energy of the relative
motion, and $W_{L}$ is the whole exchange kernel
\begin{equation}
 W_{L}(\boldsymbol{R'',R'})=H^{E}(\boldsymbol{R'',R'})-EN^{E}(\boldsymbol{R'',R'}),
\end{equation}
where the exchange kernels of the hamiltonian and overlap are
defined as
\begin{eqnarray}
H^E(\boldsymbol{R}'',\boldsymbol{R}') \!\! & = & \!\!
\langle\phi_1\phi_2\delta(\boldsymbol{R-R}'')|H|{\cal
A}'[\phi_1\phi_2\delta(\boldsymbol{R-R}')]
\rangle , \nonumber \\
N^E(\boldsymbol{R}'',\boldsymbol{R'}) \!\! & = & \!\!
\langle\phi_1\phi_2\delta(\boldsymbol{R-R}'')|{\cal
A}'|\phi_1\phi_2\delta(\boldsymbol{R-R'})
\rangle . \nonumber \\
\end{eqnarray}

\section{The results and discussions}

In order to study the resonance effects in the $NN$
scattering observables, we calculate the $NN$ scattering phase
shifts with and without the $\Delta\Delta$ state,
respectively. All the coupled channels are listed in Table
\ref{channels}.

\begin{table}[h]
\caption{The channels involved in the calculations.}
\begin{tabular}{ccccccccccc}
\hline \hline
  sc. & 2cc. & 3cc.  \\ \hline
 ~~~$^{3}D_{3}^{NN}$ ~~~ & ~~$^{3}D_{3}^{NN} + ^{7}S_{3}^{\Delta\Delta}$ ~~ &
 ~~ $^{3}D_{3}^{NN} + ^{3}G_{3}^{NN} + ^{7}S_{3}^{\Delta\Delta}$~~ \\
 $^{3}G_{3}^{NN}$  & $^{3}G_{3}^{NN} + ^{7}S_{3}^{\Delta\Delta}$ & \\
                   & $^{3}D_{3}^{NN} + ^{3}G_{3}^{NN}$ &   \\ \hline
\hline
\end{tabular}
\label{channels}
\end{table}

We first discuss the results from QDCSM. The results are shown in
Fig. 1. The experimental information used for the comparison is
the partial waves solution SP07~\cite{SP07}.

\begin{figure}[ht]
\epsfxsize=3.7in \epsfbox{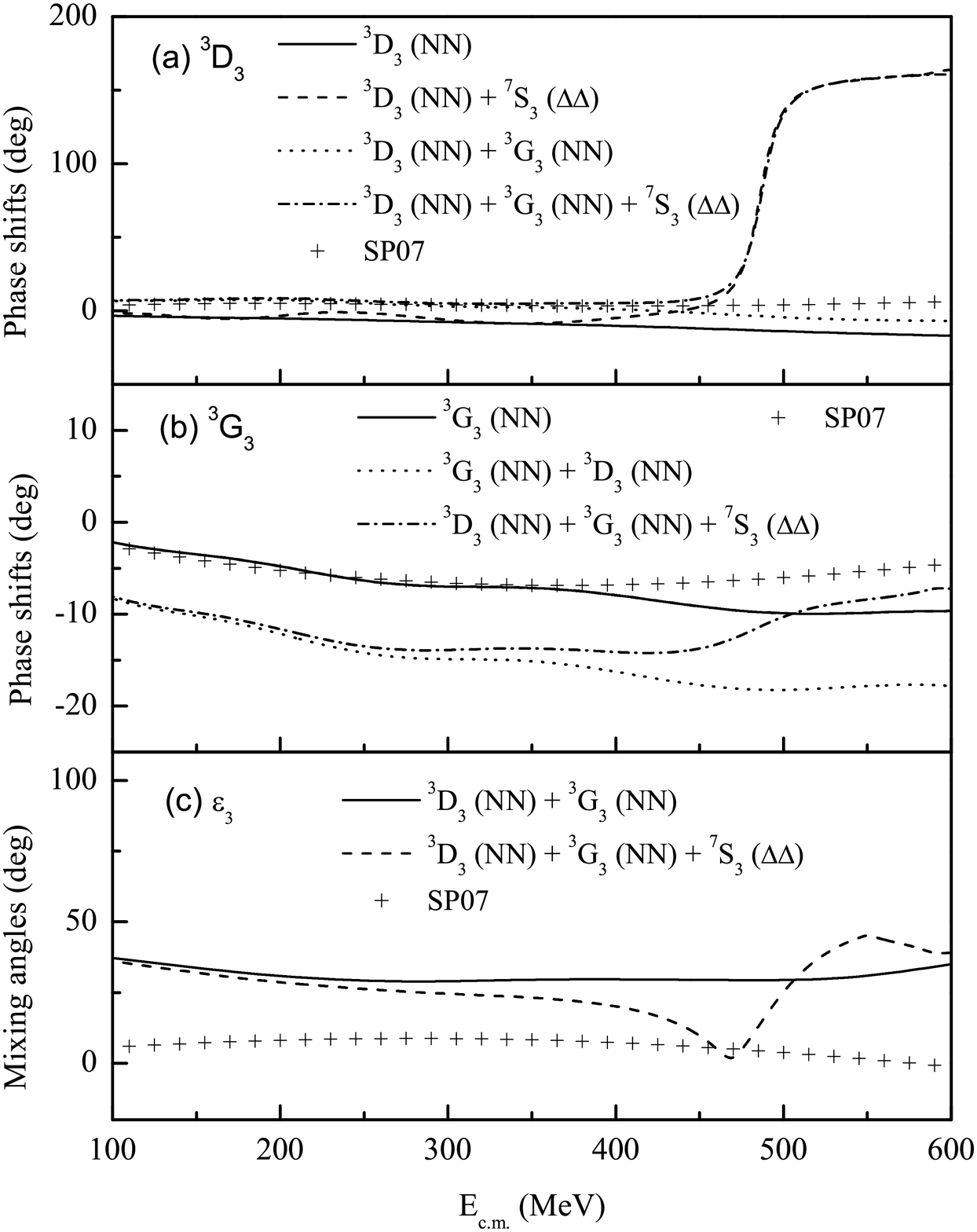} \vspace{-1.0cm}
\caption{$^{3}D_{3}^{NN}$ and $^{3}G_{3}^{NN}$ phase shifts
including their mixing angles $\varepsilon_{3}$ in QDCSM. }
\end{figure}

\begin{figure}[ht]
\epsfxsize=3.7in \epsfbox{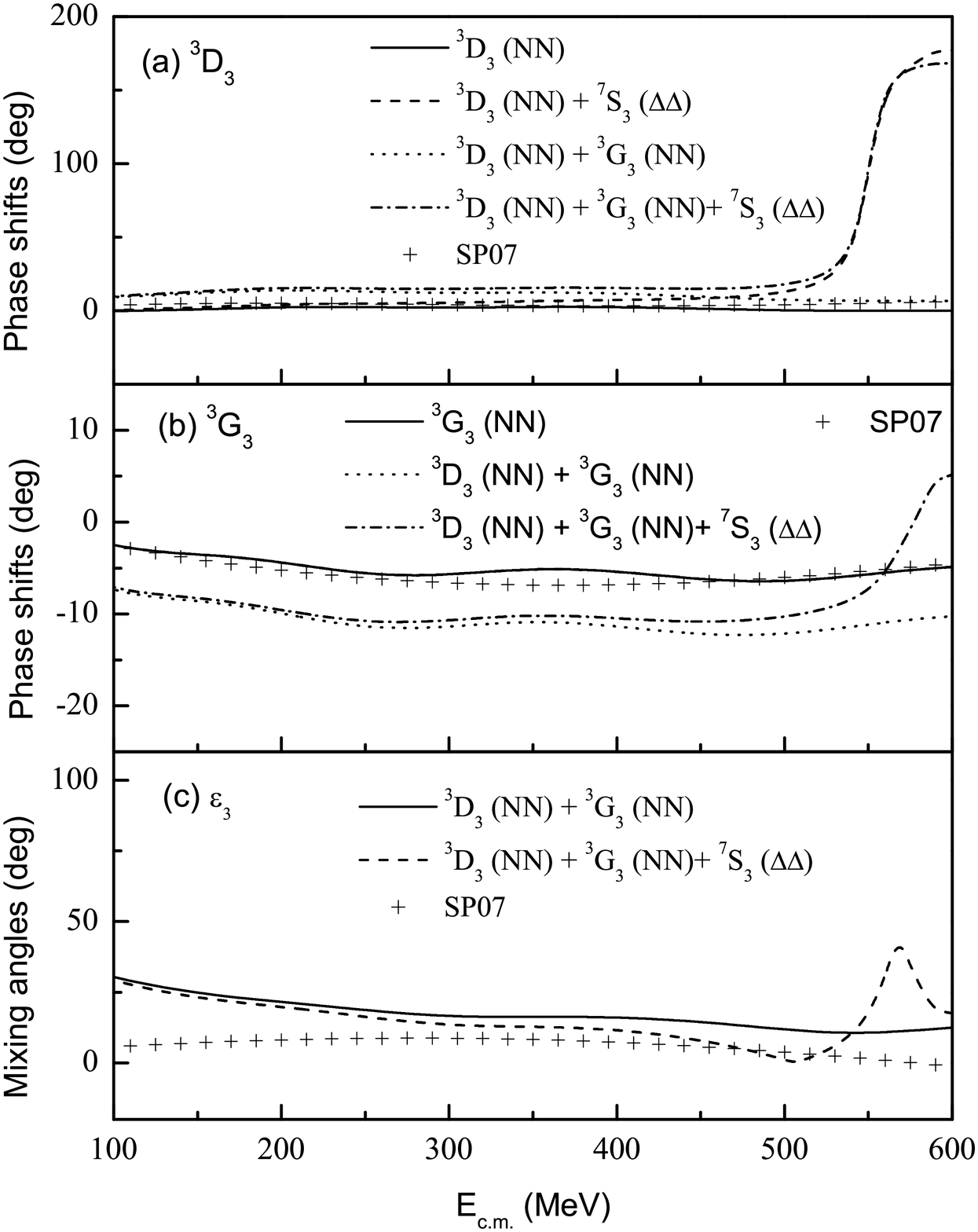} \vspace{-1.0cm}
\caption{$^{3}D_{3}^{NN}$ and $^{3}G_{3}^{NN}$ phase shifts
including their mixing angles $\varepsilon_{3}$ in ChQM. }
\end{figure}

Firstly, we do a single channel calculation for the phase shifts
of $^{3}D_{3}$ and $^{3}G_{3}$ partial waves of $NN$ scattering,
which are denoted by the solid lines in Figs. 1(a) and 1(b),
respectively. No resonance structure can be found in both two
channels as expected. Both the $^{3}D_{3}^{NN}$ and
$^{3}G_{3}^{NN}$ scattering phase shifts fit the SP07 data well,
especially for the low energy scattering, $ E_{CM} < 400$ MeV. For
$^{3}G_{3}^{NN}$ partial wave, the phase shifts at high energy, $
E_{CM} > 400$, deviate from the SP07 data, which rise up slowly.

Secondly, two-channel coupling calculations are performed. The state
$^{7}S_{3}^{\Delta\Delta}$ is added in the phase shift calculations of
the $^{3}D_{3}^{NN}$ and $^{3}G_{3}^{NN}$ partial waves of $NN$ scattering.
The results are shown in Fig. (1a) by the dashed line.
From Fig. 1(a) we can see that the channel coupling of $^{7}S_{3}^{\Delta\Delta}$
has minor effects on phase shifts of $^{3}D_{3}^{NN}$ for $ E_{CM} < 400$ MeV, but
it causes that the phase shifts rise through $\pi/2$ at a resonance mass
$\sim 2360$ MeV. For $^{3}G_{3}^{NN}$ phase shifts, there is no coupling between
$^3G^{NN}_3$ and $^{7}S_{3}^{\Delta\Delta}$, so the channel coupling phase shifts is
exactly the same as the single channel ones. The reason for this is that,
if only the two-body interactions are used, as our two quark models assumed,
the two-body tensor interaction can only induce a $\Delta{L}=2$ change at a time.
Therefore, the $^{7}S_{3}^{\Delta\Delta}$ cannot couple to $^{3}G_{3}^{NN}$ directly,
and the $^7S^{\Delta\Delta}_3$ state must go through the intermediate $^{3}D_{3}^{NN}$
channel to affect the $^3G^{NN}_3$ channel. To verify this point, we do the
following calculations.

Thirdly, we calculate the phase shifts of the coupled
$^{3}D_{3}^{NN}$ and $^{3}G_{3}^{NN}$ partial waves. The phase
shifts of $^{3}D_{3}^{NN}$ and $^{3}G_{3}^{NN}$ scattering are
denoted by the dotted lines in Fig. 1(a) and Fig.1(b), respectively.
Obviously, there is no resonance structure in any
channel as expected. By coupling to $^{3}G_{3}^{NN}$, the
phase shifts of $^{3}D_{3}^{NN}$ fit the SP07 data
better than the pure $^{3}D_{3}^{NN}$ does. However, the coupling to
$^{3}D_{3}^{NN}$ push down the phase shifts of $^{3}G_{3}^{NN}$
from the SP07 data a little.

Fourthly, we include the $^{7}S_{3}^{\Delta\Delta}$ state in the phase shift
calculation of coupled $^{3}D_{3}^{NN}$ and $^{3}G_{3}^{NN}$ partial waves.
The phase shifts of $^{3}D_{3}^{NN}$ and $^{3}G_{3}^{NN}$
are denoted by the dash-dotted lines in Figs. 1(a) and 1(b), respectively.
We find that there is a remarkable
resonance structure in the $^{3}D_{3}^{NN}$ partial wave, whereas in
the $^{3}G_{3}^{NN}$ phase shifts, there is only a small bump around
the resonance energy. This result is consistent with the recent
observations of the WASA-at-COSY Collaboration, in which
$^{3}D_{3}^{NN}$ wave obtained a typical resonance shape,
whereas the $^{3}G_{3}^{NN}$ wave changed less
dramatically~\cite{ABC5}. In addition, we can see the effect
of $^{7}S_{3}^{\Delta\Delta}$ on the phase shifts of $^{3}G_{3}^{NN}$
clearly. It proceeds through the intermediate channel $^{3}D_{3}^{NN}$.

Finally, we give the mixing angles of the coupled $^{3}D_{3}^{NN}$
and $^{3}G_{3}^{NN}$ partial waves in Fig. 1(c). The solid
(dashed) curves denote the result without (with) the resonance
state $^{7}S_{3}^{\Delta\Delta}$ in the coupled $^{3}D_{3}^{NN}$
and $^{3}G_{3}^{NN}$ partial waves. A valley and a peak
around the resonance energy appear by including the resonance state
$^{7}S_{3}^{\Delta\Delta}$.

Then we do the same calculations by using another quark model ChQM.
The results are shown in Fig. 2. The meaning of the curves
are the same as those in Fig. 1. From Fig. 2, we find that all the
results are consistent with the ones from QDCSM. Only the resonance
mass moves to $\sim 2410$ MeV, which is higher than the observed one
and that in QDCSM by $\sim 30(50)$ MeV.

In order to compare our results with the recent observations of
the WASA-at-COSY Collaboration~\cite{ABC5}, we show the
$^{3}D_{3}^{NN}$ and $^{3}G_{3}^{NN}$ amplitudes including their
mixing amplitude $\varepsilon_{3}$ for the two quark models in
Fig. 3. From Fig. 3(a), we can see that the $^{3}D_{3}^{NN}$ partial
wave obtains a typical resonance structure in both models, which is consistent with
Ref.~\cite{ABC5}. But the
resonance mass is $\sim 2360$ MeV in QDCSM, $\sim 2410$ MeV in
ChQM, a little lower and higher than $\sim 2380$ MeV in Ref.~\cite{ABC5}.
From Fig. 3(b), we
can see that both quark models give similar amplitudes of the
$^{3}G_{3}^{NN}$ partial wave. Although there are small structures
around the resonance energy, the theoretical results
are different from the experimental ones. The real part of amplitude
in model calculation rises a little after the resonance energy,
whereas the experimental ones fall down. For imaginary part, we have a
total inversed results. The amplitude falls down a little in theoretical
calculation and rises in experimental measurement~\cite{ABC5}.
For the mixing amplitude $\varepsilon_{3}$ as
shown in Fig. 3(c), the real parts of the partial wave in two
quark models are consistent with that of Ref.~\cite{ABC5}, there
is a valley around the resonance energy, whereas the imaginary
parts in both models are different from that of Ref.~\cite{ABC5}.

\begin{figure}[ht]
\epsfxsize=3.4in \epsfbox{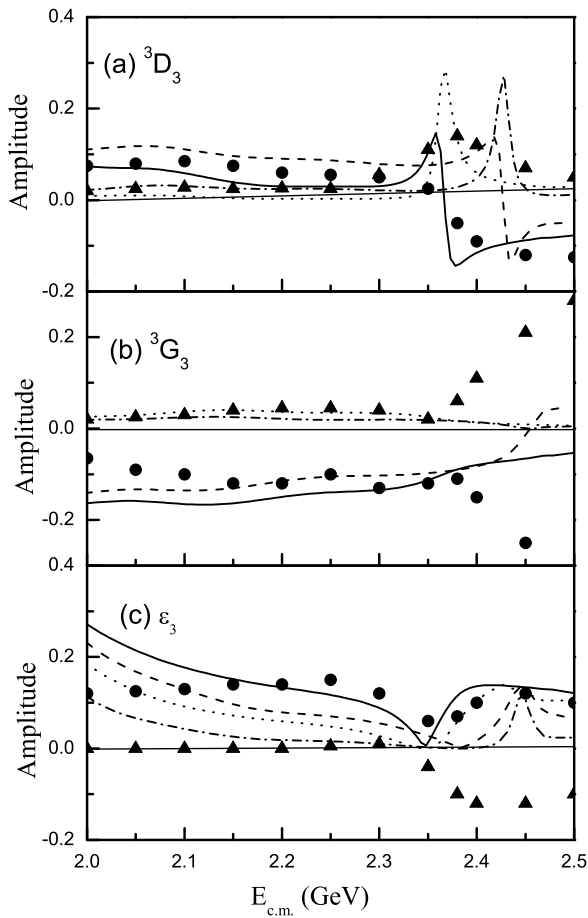} \vspace{-0.2cm}
\caption{$^{3}D_{3}^{NN}$ and $^{3}G_{3}^{NN}$ amplitudes
including their mixing amplitude $\varepsilon_{3}$ in two quark
models. Solid (dotted) curves give the real (imaginary) part of
partial-wave amplitudes in QDCSM, whereas the dashed (dash-dotted)
curves represent the real (imaginary) part of partial wave
amplitudes in ChQM. Results from Ref.~\cite{ABC5} are shown as
solid circles (real part) and solid triangles (imaginary part).}
\end{figure}

\section{Summary}

In conclusion, inspired by the recent results of the
WASA-at-COSY Collaboration, where they did a
partial wave analysis including the new $np$ scattering data
and found a resonance pole in the coupled $^{3}D_{3}$ and
$^{3}G_{3}$ partial waves as expected from the $d^{*}$ resonance
hypothesis, we study the $d^{*}$ resonance structure in the
coupled $^{3}D_{3}$ - $^{3}G_{3}$ partial waves of $NN$ scattering
in the framework of QDCSM and ChQM. Our results show that there
is indeed a resonance in the coupled $^{3}D_{3}^{NN}$ - $^{3}G_{3}^{NN}$
partial waves due to
the coupling of the $^{7}S_{3}^{\Delta\Delta}$ state,
which is consistent with the existence of the $d^{*}$
resonance. The resonance structure in the $^{3}D_{3}^{NN}$
partial wave is remarkable, whereas in the $^{3}G_{3}^{NN}$ phase
shifts there is only a small bump around the resonance energy. This
result is also consistent with the recent observations of the
WASA-at-COSY Collaboration~\cite{ABC5}. From our calculation, one
can see that the small bump in the phase shifts of $^{3}G_{3}$
partial wave of $NN$ scattering is an indication of the existence
of a resonance, $^{7}S_{3}^{\Delta\Delta}$.

Moreover, QDCSM and ChQM obtained similar results. Only the
mass and decay width of the $d^{*}$ resonance in QDCSM are smaller
than that in ChQM. It shows once again the consistency of these two
quark models even though they have different intermediate-range attraction
mechanisms. Our former study has shown that by including the
hidden-color channels in ChQM, the resonance masses are lowered by
10-20 MeV~\cite{Ping_NN,Huang}. This fact inferred that the
quark delocalization and color screening used in QDCSM might be an
effective description of the hidden color channel coupling.

For the partial wave amplitudes, the behavior of the theoretical calculated ones
are different from the measured ones. The further study is needed.

\section*{Acknowledgment}
This work is supported partly by the
National Science Foundation of China under Contract Nos. 11205091, 11035006 and 11175088.

\end{document}